\documentclass[12pt]{iopart}
\usepackage{iopams,graphicx,color} 
\newcommand{\half}{{\case{1}{2}}}

\newcommand{\ez}{{\mathbf {\hat e}_{z}}}

\newcommand{\uvec}{{\mathbf u}}
\newcommand{\xvec}{{\mathbf x}}
\newcommand{\avec}{{\mathbf a}}

\newcommand{\Pvec}{{\mathbf \Pi}}
\newcommand{\kvec}{{\mathbf k}}

\newcommand{\Uvec}{{\mathbf U}}
\newcommand{\vvec}{{\mathbf v}}
\newcommand{\mvec}{{\mathbf m}}
\newcommand{\Rvec}{{\mathbf R}}
\newcommand{\Svec}{{\mathbf S}}

\newcommand{\Nmat}{{\mathcal N}}
\newcommand{\Pmat}{{\mathcal P}}
\newcommand{\Smat}{{\mathcal S}}
\newcommand{\curl}{{\rm curl}}

\newcommand{\Ovec}{{\mathbf \Omega}}
\newcommand{\zerovec}{{\mathbf 0}}

\newcommand{\fd}[1]{\frac{\mathrm{d}{#1}}{\mathrm{d}t}}
\newcommand{\dotp}[2]{  {#1}\cdot {#2}  }
\newcommand{\newt}[1]{\textcolor{blue}{#1}}
\newcommand{\remtext}[1]{}
\newcommand{\remfigure}[1]{#1}
\begin{document}
\title[EP formulation of $n^{th}-$gradient fluids]{Euler-Poincar\'e
   formulation and elliptic instability 
   for $n^{th}-$gradient fluids}
\author{Bruce R. Fabijonas\dag\  
and Darryl D. Holm\ddag\ \S\ 
}

\address{\dag\ Department of Mathematics,
      Southern Methodist University,
      Dallas, TX 75275-0156, USA, email: bfabi@smu.edu}

\address{\ddag\ Theoretical Division and Center for Nonlinear Studies,
      Los Alamos National Laboratory,
      Los Alamos, NM 87545, USA, email: dholm@lanl.gov}

\address{\S\ Mathematics Department,
      Imperial College London,
      SW7 2AZ, UK, email: d.holm@imperial.ac.uk}

\begin{abstract}
The energy of an $n^{th}-$gradient fluid depends on its Eulerian velocity
gradients of order $n$.  A variational principle is introduced for
the dynamics of $n^{th}-$gradient fluids and their properties are reviewed
in the context of Noether's theorem. The stability properties of
Craik-Criminale solutions for first and second gradient fluids are
examined.
\end{abstract}

\submitto{\JPA}
\section{Introduction}\label{Intro-sec}

Classical continuum theories lack any length scales and as such provide
leading order approximations for a number of problems that contain
microstructures. Microstructures typically introduce characteristic length
scales that may induce gradient dependences of various kinds. Several
continuum theories have been developed to deal with microstructures and
their attendant phenomena. These include micropolar, micromorphic,
strain-gradient, non-local, etc. See Eringen \cite{Eringen[1976]} and Nowacki
\cite{Nowacki[1986]}  for
catalogs of such phenomenological theories. Physical theories for complex
fluids such as liquid crystals have also been introduced, based on symmetry
breaking phase transitions that yield statistically-defined order
parameters as additional thermodynamic variables.   See de Gennes and
Prost \cite{deGennesProst[1993]} 
for discussions of the fundamental principles of order
parameter physics for liquid crystals. See Holm \cite{holm:02} for a
variational 
description of order parameter theories of complex fluids. Often these
phenomenological theories are combined with geometrical discussions based
on the theory of Cosserat and Cosserat \cite{Cosserats[1909]}.

The mathematical theory of continuum mechanics for complex, or composite
materials produced a number of interesting phenomenological models in the
1960's. Among these are the models of differential type introduced by
Rivlin and Ericksen \cite{rivlin:erickson:55} and the multipolar 
models of Green and Rivlin
\cite{green:rivlin:64a,green:rivlin:64b}. 
The history of how these models were tested in comparison with
experiments and refined by {\it ab initio} assumptions of thermodynamics is
recounted, for example, in Dunn and Fosdick \cite{Dunn-Fosdick[1974]},
Eringen \cite{Eringen[1976]} and 
Fosdick and Rajagopal \cite{Rajagopal[1980s]}.

Recently, Bellout \etal \cite{Bellout-etal[1999]} considered a
fusion of the models due 
to Rivlin and Ericksen \cite{rivlin:erickson:55} and those of Green and Rivlin
\cite{green:rivlin:64a,green:rivlin:64b}.  The
resulting theory introduced higher order spatial velocity gradients into
the energy that regularized the model solutions and endowed the model with
promising stability characteristics.

The present work specializes to a subclass of the Rivlin-Ericksen-Green
multipolar fluids treated in Bellout \etal
\cite{Bellout-etal[1999]}  that has energy
density given by
\begin{eqnarray}\label{grad-n-erg}
\half D|\uvec|^2
+ DW(\mathbf{e},\nabla\mathbf{e},\nabla\nabla\mathbf{e},
\dots,D,\nabla{D},\nabla\nabla{D},\dots),
\end{eqnarray}
where $D$ is the mass density and
$\mathbf{e}=\half(\nabla\uvec+\nabla\uvec^T)$
is the strain rate tensor. Materials whose energy density takes this form
are called {\it gradient fluids of degree $n$}, where $n$ is the order
of the velocity gradients appearing in (\ref{grad-n-erg}). The case $n=0$
is the Euler fluid (no velocity gradient dependence), while the case $n=1$
coincides with the $2^{nd}$ grade fluid \cite{Dunn-Fosdick[1974]},
whose energy depends on the 
velocity gradient through the strain-rate, $\mathbf{e}$.

\newt{%
Our aim here is to investigate the implications of adopting a subclass 
of these $n^{th}-$gradient models for the well-known elliptic instability, 
which governs the rapid, violent transition from two-dimensional to 
three-dimensional motion at the onset of turbulence in Newtonian 
fluids \cite{pierrehumbert:86,bayly:86,land:saff:87,wal:90,kers:02}.
We shall not assess the implications for experimental measurements of 
this investigation, as we feel that such an assessment may still be 
premature. Instead, we continue the investigation begun by 
Bellout \etal \cite{Bellout-etal[1999]} in studying the role 
of $n^{th}-$gradient constitutive 
relations on fluid instability. We begin by casting the $n^{th}-$gradient 
theory of nonlinear elasticity into the Euler-Poincar\'e variational 
framework \cite{holm:mar:rat:98b}. 
The Euler-Poincar\'e framework allows us to take advantage 
of several parallels 
between $n^{th}-$gradient fluids and recently developed 
Lagrangian-averaged Navier-Stokes-alpha, or LANS$-\alpha$, 
turbulence closure models of Foias \etal \cite{foias:holm:titi:01}.
Since Rivlin \cite{rivlin:57}, 
remarkable parallels have been drawn between nonlinear elasticity and 
turbulence closure models. In our case, the Euler-Poincar\'e framework 
leads to energy 
balance laws, a proper definition of momentum density, circulation 
theorems and to the Craik-Criminale (CC) class of exact 
solutions for the $n^{th}-$gradient materials. }

\newt{%
The CC solutions \cite{craik:crim:86} form the basis for 
analyzing elliptic instability, in 
which two-dimensional flows with closed streamlines are subject 
to three-dimensional
instabilities. Our aim in this paper
is to determine the effects of $n^{th}-$gradient 
viscoelasticity on the parametric resonance mechanism responsible for 
elliptic instability and on its growth rates. We follow the earlier 
treatment of elliptic instability for Newtonian fluids as reviewd, 
e.g., by Kerswell \cite{kers:02}, and we are guided 
by the results of Fabijonas and Holm \cite{fab:holm:03a,fab:holm:03c} based 
on the CC solutions for 
the LANS$-\alpha$ and similar closure models for turbulence. 
Thus, we consider plane wave disturbances of 
elliptical flows whose wave amplitude and wave vector are time-dependent. 
This approach leads to a Floquet problem for the wave amplitude of the
disturbance. Remarkably, we discover that these viscoelastic 
effects may be either stabilizing, or destabilizing, in the sense 
that they alter the shape and size of the 
instability domain while simultaneously 
increasing or decreasing the associated Lyapunov growth rates,
depending on the parameter values. We hope
that experimentalists  
may be guided by these results in testing whether $n^{th}-$gradient models 
may be appropriate for the description of viscoelastic materials undergoing 
elliptic instability.}

The equations of motion for gradient fluids are obtained from the Eulerian
form of Hamilton's principle introduced in Holm, Marsden and Ratiu
\cite{holm:mar:rat:98b} 
called the Euler-Poincar\'e theory for continua with advected quantities.
In vector notation, this is
\begin{eqnarray}
\frac{\partial}{\partial t} \mvec
+
\uvec\cdot\nabla\mvec
+
(\nabla \uvec)^T\cdot\mvec
+
\mvec {\rm\,div\,}\uvec
- \nabla\frac{\delta {L}}{\delta D} = 0
\,,\,\,\hbox{where}\,\,
\mvec \equiv  \frac{\delta {L}}{\delta\uvec}
\,.
\label{EP-motion}
\end{eqnarray}
For the class of Lagrangians we shall consider, for $0^{th}$,
$1^{st}$ and $2^{nd}$ gradient fluids, one has
\begin{eqnarray}
{L} = \int d^3x \Big \{ \half D|\uvec|^2
+ D\uvec\cdot\Rvec(\xvec) - p(D-1)
+ DW(\mathbf{e},\nabla\mathbf{e}) \Big \}
\,.\label{Lag-Grade-2-3}
\end{eqnarray}
The term in $\Rvec$ in this Lagrangian boosts the gradient fluid flow into
a frame rotating with angular frequency $2\Ovec = \curl\,\Rvec$, while the
term in $p$ imposes the constraint $D=1$. Hence $\nabla\cdot\uvec=0$, as implied
by substituting $D=1$ into the continuity equation,
\begin{eqnarray}
\partial_tD+\nabla\cdot( D\uvec) =0
\,.\label{cont-eqn}
\end{eqnarray}
Many mathematical regularity properties are available for the class of
gradient fluids, especially for the case that the Lagrangian $L$ in
(\ref{Lag-Grade-2-3}) provides a norm (when evaluated on the constraint
surface, $D=1$). However, these regularity properties for gradient
fluids will be discussed elsewhere, following Foias \etal
\cite{Foias[2002]}. 

The objective of the current paper is to investigate the stability
properties of CC solutions of the gradient fluid
equations. For CC solutions in an unbounded domain, the fluid velocity is
linear in the spatial coordinate and the pressure is quadratic. The CC
solutions may be regarded as the first term in a Taylor expansion in
space, around a stagnation point of the gradient fluid flow in a moving
frame.  We shall use the theory of elliptic instability to investigate
the exact nonlinear growth rates when CC solutions interact with a wave
packet whose phase is frozen into the CC flow for gradient fluids of
degree $n=1,2$. (The Euler case $n=0$ was studied in the original work
of Craik and Criminale \cite{craik:crim:86}. See also Craik
\cite{craik:89}, Miyazaki \cite{miya:93} and 
Kerswell \cite{kers:02} for subsequent developments.  See also Lagnado
and Simmen \cite{lag:simm:93} and Goddard and Alam \cite{god:alam:99}
for similar analyses for an upper-convected Maxwell fluid and
granular media, respectively.)

\paragraph*{Outline.} Section \ref{EP-form-sec} summarizes the properties of
ideal gradient fluids that follow directly from their Euler-Poincar\'e
formulation. These properties include energy conservation, momentum balance
and Kelvin circulation preservation, all of which follow from Noether's
theorem. We then specialize to gradient fluids of degree $n=1,2$. Section
\ref{CC-soln-sec} introduces the CC solutions for $1^{st}$ and $2^{nd}$
gradient fluids. Section \ref{Elliptic-Instab-sec} discusses their stability
properties for both inviscid and viscous CC solutions. Here we introduce
viscosity as in the theory of $2^{nd}$ grade fluids, to which the gradient
fluids reduce when $n=1$. Section \ref{conclusion-sec} summarizes our
conclusions.

\section{EP formulation of gradient fluids}\label{EP-form-sec}

\paragraph*{Hamilton's principle for first and second gradient fluids.}
The mathematical basis common to all ideal fluid motions is Hamilton's
principle
\begin{eqnarray}
\delta \int {L}~dt = 0,
\end{eqnarray}
where ${L}$ is the Lagrangian for the system.    We work in the Eulerian
representation of fluids, where the Euler-Lagrange equation is
replaced by the Euler-Poincar\'e equation.  See
Ref.~\cite{holm:mar:rat:98b} for a detailed discussion of Euler-Poincar\'e
theory.

This paper focuses on the incompressible motion of first and second 
gradient fluids in a rotating frame.  Thus, the class of Lagrangians 
we
shall consider  has the form \cite{holm:mar:rat:98b}
\begin{eqnarray}
{L}
= \int
\mathcal{L}({\mathbf{u}},\nabla{\mathbf{u}},
\nabla\nabla{\mathbf{u}},\dots,D,\nabla{D},
\nabla\nabla{D},\dots;\Rvec(\xvec))\,d^3x
\,,
\label{Lag-multipolar}
\end{eqnarray}
where $\Rvec(\xvec)$ is the vector potential for
the Coriolis parameter, i.e., $\curl\,\Rvec = 2\Ovec$.
Specifically, we shall take,
\begin{eqnarray}
{L} = \int d^3x \Big \{ \half D|\uvec|^2
+ D\uvec\cdot\Rvec(\xvec) - p(D-1)
+ DW(\mathbf{e},\nabla\mathbf{e}) \Big \}
\,.\label{Lag-Grade2-3}
\end{eqnarray}
Here $p$ is pressure (a Lagrange multiplier), $D$ is mass density,
and $\uvec$ is fluid velocity. Through the function
$W(\mathbf{e},\nabla\mathbf{e})$, the first and second gradient fluids
depend on $\mathbf{e}$, the symmetric strain-rate tensor,
\begin{eqnarray}
e_{ij} = \half(u_{i,j} + u_{j,i} ) = \frac{1}{2}\left ( \frac{\partial
u_i}{\partial x^j} + \frac{\partial u_j}{\partial x^i} \right )
\,.
\end{eqnarray}
That is, first and second gradient fluids allow energy to
depend upon strain-rate $\mathbf{e}$ and gradient of strain-rate
$\nabla\mathbf{e}$, respectively,
\cite{Eringen[1976],Dunn-Fosdick[1974],Rajagopal[1980s]}.
The higher gradient fluids will allow energies that
depend upon higher-order gradients of strain-rate. We introduced the
dependence on $\mathbf{e}$, $\nabla\mathbf{e}$, etc., instead of
$\nabla\mathbf{u}$, $\nabla\nabla\mathbf{u}$, etc., in equation
(\ref{Lag-Grade2-3}), so that the Lagrangian $L$ will be invariant under
rotations. Consequently, the resulting Euler-Poincar\'e equations will
admit an angular momentum balance relation and will satisfy the
requirements of material frame indifference.

\paragraph*{Variational derivatives and natural boundary conditions.}
The variational derivatives of the Lagrangian (\ref{Lag-Grade2-3}) for
first and second gradient fluids are given by
\begin{eqnarray}
\fl \delta {L} = \int d^3x \Big \{ D(\uvec+\Rvec)\cdot\delta\uvec +
    \left ( \half |\uvec|^2+\uvec\cdot\Rvec+W-p \right ) \delta D
    - (D-1)\delta p  + \boldsymbol{\sigma:\,}\delta\mathbf{e}\Big \},
\end{eqnarray}
where
$\boldsymbol{\sigma:\,}\delta\mathbf{e}
={\rm\,tr}(\boldsymbol{\sigma^T\cdot\,}\delta\mathbf{e})
=\sigma^{ij}\delta{e}_{ij}$, and we sum over
repeated indices. The quantity $\boldsymbol{\sigma}$ is the stress tensor,
whose definition assures that it is  symmetric,
$\boldsymbol{\sigma^T}=\boldsymbol{\sigma}$,
\begin{eqnarray}
\sigma^{ij} \equiv \frac{\delta {L}}{\delta e_{ij}}
     =
     D\,\frac{\partial W}{\partial e_{ij}}
- \nabla\cdot D\,
\frac{\partial W}{\partial\nabla e_{ij}}
\,.\label{stress-tensor}
\end{eqnarray}
The added natural boundary condition for second gradient fluids,
\begin{eqnarray}
{\mathbf {\hat n}}\cdot\frac{\partial W}{\partial\nabla e_{ij}}\delta
e_{ij} = 0
\,,
\end{eqnarray}
arises from an integration by parts.  Another application
of integration by parts and use of the symmetry of $e_{ij}$ gives
\begin{eqnarray} \label{Lagrangn}
\fl \delta {L} = \int d^3x \Big
\{ D(\uvec+\Rvec-\nabla\cdot{\boldsymbol\sigma})\cdot\delta\uvec
+ (\half|\uvec|^2+\uvec\cdot\Rvec+W-p)\delta
D  - (D-1)\delta
p  \Big \},
\end{eqnarray}
where $(\nabla\cdot{\boldsymbol\sigma})^i = \partial\sigma^{ij}/\partial
x^j$. Another natural boundary condition has been introduced and applied,
\begin{eqnarray}
{\mathbf {\hat n}}\cdot{\boldsymbol\sigma}\cdot\delta\uvec = 0
\,,\quad\hbox{at the boundary.}
\end{eqnarray}
This condition may be satisfied when the fluid velocity has no normal
component at the boundary, by requiring that the normal stress have no
tangential component,
\begin{eqnarray}
({\mathbf {\hat n}}\cdot{\boldsymbol\sigma})
\times{\mathbf {\hat n}}=0
\,,\quad\hbox{at the boundary.}
\label{normal-stress-bc}
\end{eqnarray}

\paragraph*{The Euler-Poincar\'e motion equation.}
The Euler-Poincar\'e motion equation is
\cite{holm:mar:rat:98b}
\begin{eqnarray}
\fl \frac{\partial}{\partial t} \mvec
+
\uvec\cdot\nabla\mvec
+
(\nabla \uvec)^T\cdot\mvec
+
\mvec {\rm\,div\,}\uvec
- \nabla\frac{\delta {L}}{\delta D} = 0
\,,\,\,\hbox{where}\,\,
\mvec \equiv  \frac{\delta {L}}{\delta\uvec}
\,.
\label{EPmotion}
\end{eqnarray}
The momentum density $\mvec$ is defined as the variational
derivative of the Lagrangian with respect to the fluid velocity $\uvec$.
For the gradient fluid Lagrangian (\ref{Lag-Grade2-3}), we see from
\eref{Lagrangn} that this is
\begin{eqnarray}
\mvec \equiv  \frac{\delta {L}}{\delta\uvec}
= D(\uvec+\Rvec-\nabla\cdot{\boldsymbol\sigma})
\,.
\end{eqnarray}
We denote $(\nabla \uvec)^T\cdot\mvec=m_j\nabla{u}^j$, and
$d/dt = \partial/\partial t + \uvec\cdot\nabla$
is the material derivative along $\uvec$. The incompressibility condition
$\nabla\cdot\uvec = 0$  follows from the continuity
equation $\partial_tD+\,\nabla\cdot\,(D{\mathbf{u}})=0$, evaluated for $D=1$,
as imposed by the pressure constraint. Consequently,  the Euler-Poincar\'e
motion equation (\ref{EPmotion}) obtained from the gradient fluid Lagrangian
(\ref{Lag-Grade2-3}) is expressed as,
\begin{eqnarray}
\fl \frac{d}{dt}\left ( \uvec+\Rvec-\nabla\cdot{\boldsymbol\sigma} \right )
+
(\nabla\uvec)^T \cdot
\left ( \uvec+\Rvec-\nabla\cdot{\boldsymbol\sigma} \right )+
   \nabla \left ( p - \half |\uvec|^2 - W - \uvec\cdot\Rvec\right )
= 0
\,,
\end{eqnarray}
together with $\nabla\cdot\uvec = 0$.
Next, we use the vector identity
\[(\uvec\cdot\nabla)\Rvec
+ (\nabla\uvec)^T\cdot\Rvec
= -\,\uvec\times\curl\,\Rvec
+ \nabla(\uvec\cdot\Rvec)
\,,\]
together with the Coriolis relation $\curl\,\Rvec=2\Ovec(\xvec)$,
and introduce the standard dissipation law for the first and second 
gradient fluids. Consequently, the motion equation takes the familiar 
form,
\begin{eqnarray}
\fl \frac{d}{dt}\left ( \uvec-\nabla\cdot{\boldsymbol\sigma} \right )
+
(\nabla\uvec)^T \cdot\left ( \uvec-\nabla\cdot{\boldsymbol\sigma} \right )
+
2\Ovec\times\uvec
+
   \nabla \left ( p - \half |\uvec|^2 - W \right )= \nu\Delta\uvec.
\label{motion-eqn}
\end{eqnarray}

\subsection{Circulation theorem and energy-momentum conservation}
\paragraph*{Kelvin-Noether circulation theorem.} In the absence of forcing
and dissipation, the Euler-Poincar\'e theory for Lagrangians in the class
(\ref{Lag-multipolar}) provides a Kelvin-Noether circulation theorem
\cite{holm:mar:rat:98b}
\begin{eqnarray}
\frac{d}{dt} \oint_{c(\uvec)}
\frac{1}{D} \frac{\delta {L}}{\delta\uvec}
\cdot d\xvec
= 0
\,,
\end{eqnarray}
which holds for integrations around any closed curve $c(\uvec)$ moving
with the fluid. For the first and second gradient fluids considered here,
this becomes,
\begin{eqnarray}
\frac{d}{dt} \oint_{c(\uvec)}
\Big( \uvec-\nabla\cdot{\boldsymbol\sigma}+\Rvec(\xvec)\Big)
\cdot d\xvec
= 0
\,.
\end{eqnarray}
Stokes theorem then provides, for relative vorticity
$\boldsymbol{\omega}=\curl\,\uvec$, that
\begin{eqnarray}
\frac{d}{dt} \int_{S(t)}
\Big(\boldsymbol{\omega}
-\curl(\nabla\cdot{\boldsymbol\sigma})+2\Ovec(\xvec)\Big)
\cdot d\Svec
= 0
\,,
\end{eqnarray}
for any surface $S(t)$ whose boundary $\partial{S}(\uvec)$ moves
with the fluid. Consequently, we find the Helmholtz vortex dynamics
equation for the total vorticity, in the absence of forcing and
dissipation, as
\begin{eqnarray}
\frac{\partial}{\partial t} \boldsymbol{\Sigma}
+ \uvec\cdot\nabla\,\boldsymbol{\Sigma}
-\boldsymbol{\Sigma}\cdot\nabla\uvec
= 0
\,,\,\,\hbox{where}\,\,
\boldsymbol{\Sigma}
= \boldsymbol{\omega}-\curl(\nabla\cdot{\boldsymbol\sigma})+2\Ovec(\xvec)
\,.\label{Helm-vort-dyn}
\end{eqnarray}
Thus, the Kelvin-Noether circulation theorem in the Euler-Poincar\'e
framework implies that the total vorticity $\boldsymbol{\Sigma}$ is
frozen into the flow of a non-Newtonian, first or second gradient
fluid. Hence, its total vorticity $\boldsymbol{\Sigma}$ satisfies the
Helmholtz vortex dynamics equation (\ref{Helm-vort-dyn}).

\paragraph*{Energy conservation.}
  From the Euler-Poincar\'e theory, one may compute the Hamiltonian
from the Lagrangian ${L}$ in equation (\ref{Lag-Grade2-3}) for first 
and second gradient fluids in a rotating frame by applying the 
Legendre
transformation,%
\footnote[4]{Actually, we compute only the Routhian; because we do not
Legendre transform the pressure; and we do not complete the transformation
to explicit dependence only on $\mvec$.}
\begin{eqnarray}
\fl H = \langle\mvec,\uvec\rangle - {L}
\nonumber \\
\lo=
\int\!\!\int\!\!d^3x \Big \{ \half D|\uvec|^2
+ {\boldsymbol \sigma:\,}\mathbf{e}
- DW(\mathbf{e},\nabla\mathbf{e})
+  p(D-1) \Big \}
-\oint
{\mathbf {\hat n}}\cdot{\boldsymbol\sigma}\cdot\uvec\,dS
\,.
\end{eqnarray}
The corresponding conserved energy is found by evaluating this expression
on the constraint manifold, $D=1$, as
\begin{eqnarray}
E =
\int \int d^3x \Big \{ \half |\uvec|^2
+ {\boldsymbol \sigma:\,}\mathbf{e}
- W(\mathbf{e},\nabla\mathbf{e})
\Big \}
-\oint {\mathbf {\hat n}}\cdot{\boldsymbol\sigma}\cdot\uvec\,dS
\,.
\end{eqnarray}
The surface integrals in the last two equations vanish, upon applying the
normal-stress boundary condition (\ref{normal-stress-bc}), for the
situation in which the velocity $\uvec$ on the surface has no normal
component. As a consequence, the inner product of the fluid velocity
$\uvec$ with the motion equation (\ref{motion-eqn}) yields,
\begin{eqnarray}
\frac{dE}{dt}=-\,\nu\!\!\int |\nabla\uvec|^2\,d\,^3x
+\frac{\nu}{2}\oint \mathbf{\hat{n}}\cdot\nabla|\uvec|^2\,dS
\,.
\end{eqnarray}
The surface integral vanishes, in this energy balance relation for
first and second gradient fluids in a rotating frame, provided $\uvec$
vanishes on the boundary.

\paragraph*{Momentum conservation.}
We express the Euler-Poincar\'e equation (\ref{EPmotion}) in components as
\begin{equation}\label{EP-eqn}
\frac{\partial}{\partial t}
\frac{\delta {L}}{\delta {u}^i}
+\,
\frac{\partial}{\partial x^j}
\Big(\frac{\delta {L}}{\delta {u}^i}
{u}^j\Big)
+\,
\frac{\delta {L}}{\delta {u}^{\,j}}
\partial_i
{u}^{\,j}
-
{D}\partial_i
\frac{\delta {L}}{\delta {D}}
=
0
\,.
\end{equation}
Observe that for a gradient fluid Lagrangian (\ref{Lag-multipolar}) given by
\[
{L}
= \int
\mathcal{L}({\mathbf{u}},\nabla{\mathbf{u}},
\nabla\nabla{\mathbf{u}},\dots,D,\nabla{D},
\nabla\nabla{D},\dots)\,d^3x
\,,\]
we have variational derivatives
\begin{eqnarray}
\frac{\delta{L}}{\delta {D}}
&=&
\frac{\partial\mathcal{L}}{\partial {D}}
-
\frac{\partial }{\partial  x^l}
\frac{\partial\mathcal{L}}{\partial {D}_{,l}}
+
\frac{\partial^2 }{\partial  x^l\partial  x^m}
\frac{\partial\mathcal{L}}{\partial {D}_{,lm}}
-+\cdots
\nonumber\\
\frac{\delta {L}}{\delta {u}^i}
&=&
\frac{\partial \mathcal{L}}{\partial {u}^i}
-
\frac{\partial}{\partial x^j}
\frac{\partial \mathcal{L}}{\partial {u}^i_{\,,j}}
+
\frac{\partial^2}{\partial x^j\partial x^l}
\frac{\partial \mathcal{L}}{\partial {u}^i_{\,,jl}}
-+\cdots
\label{gen-momentum}
\end{eqnarray}
where the $-+\cdots$ refer to any dependence of the
Lagrangian density $\mathcal{L}$ on higher spatial derivatives of
$D$ and ${\mathbf{u}}$.
Therefore, upon performing the indicated differentiations by parts, one
eventually finds  the local conservation law for momentum,
\begin{equation}\label{mom-form}
\partial_t\,{m}_i
=
-\ \frac{\partial }{\partial  x^j}T\,^j_i
\,,\quad\hbox{with momentum density}\quad
m_i\equiv\frac{\delta {L}}{\delta {u}^i}
\,,
\end{equation}
and momentum-stress tensor $T\,^j_i$ defined by
\begin{eqnarray}
\fl T\,^j_i =
{m}_i\, {u}^j
+
\bigg(
\mathcal{L}
-  {D}\frac{\delta{L}}{\delta {D}}
\bigg)\delta^j_i  -\,
\bigg(
\frac{\delta{L}}{\delta {u}^k_{\,,j}}\,{u}^k_{\,,i}
+
\frac{\delta{L}}{\delta {u}^k_{\,,jl}}\,{u}^k_{\,,li}
+
\frac{\delta{L}}{\delta {u}^k_{\,,jlm}}\,{u}^k_{\,,lmi}
+\cdots
\bigg)
\,.\label{stress-tens}
\end{eqnarray}
Here we abbreviate, by using variational-derivative notation to denote,
\begin{eqnarray}
\frac{\delta{L}}{\delta {u}^k_{\,,j}}
&=&
\frac{\partial\mathcal{L}}{\partial {u}^k_{\,,j}}
-
\frac{\partial }{\partial  x^l}
\frac{\partial\mathcal{L}}{\partial {u}^k_{\,,jl}}
+
\frac{\partial^2 }{\partial  x^l\partial  x^m}
\frac{\partial\mathcal{L}}{\partial {u}^k_{\,,jlm}}
-+\cdots\,,
\nonumber\\
\frac{\delta{L}}{\delta {u}^k_{\,,jl}}
&=&
\frac{\partial\mathcal{L}}{\partial {u}^k_{\,,jl}}
-
\frac{\partial }{\partial  x^m}
\frac{\partial\mathcal{L}}{\partial {u}^k_{\,,jlm}}
+
\frac{\partial^2 }{\partial  x^m\partial  x^n}
\frac{\partial\mathcal{L}}{\partial {u}^k_{\,,jlmn}}
-+\cdots\,,
\label{gen-stress}\\
\frac{\delta{L}}{\delta {u}^k_{\,,jlm}}
&=&
\frac{\partial\mathcal{L}}{\partial {u}^k_{\,,jlm}}
-
\frac{\partial }{\partial  x^n}
\frac{\partial\mathcal{L}}{\partial {u}^k_{\,,jlmn}}
+
\frac{\partial^2 }{\partial  x^n\partial  x^p}
\frac{\partial\mathcal{L}}{\partial {u}^k_{\,,jlmnp}}
-+\cdots\,.
\nonumber
\end{eqnarray}
\remtext{
and the $-+\dots$ refers to the dependence of the Lagrangian
density $\mathcal{L}$ on higher spatial derivatives of
$D$ and ${\mathbf{u}}$. }

The momentum conservation form (\ref{mom-form}) is guaranteed by the
Euler-Poincar\'e equation for any choice of Lagrangian that does not
depend explicitly on the spatial coordinate.
The Coriolis vector potential $\Rvec(\xvec)$ introduces explicit spatial
dependence into the Lagrangian. Consequently, although not all components
of the momentum will be conserved, we may still write the motion equation
(\ref{motion-eqn}) as a {\bf momentum balance relation},
\begin{equation}\label{mom-balance}
\partial_t\,{m}_i
=
-\ \frac{\partial }{\partial  x^j}T\,^j_i
+\nu\Delta{u}_i
+\epsilon_{ijk}u^j2\Omega^k
\,,
\end{equation}
where the momentum-stress tensor $T\,^j_i $ is given by (\ref{stress-tens})
and $\epsilon_{ijk}$ is the completely antisymmetric tensor density, with
$\epsilon_{123}=1$.

Equations (\ref{gen-momentum}) for the momentum density $m_i$ and
(\ref{stress-tens}) for the momentum-stress tensor $T\,^j_i$ indicate how
the derivation and analysis may proceed within the Euler-Poincar\'e
framework for gradient fluids of degree three, four, five, etc.
These generalizations correspond to allowing the strain-rate $W$ in
the Lagrangian (\ref{Lag-Grade2-3}) to depend on higher gradients of the
strain-rate $\mathbf{e}$. Pursuing this direction further for $n^{th}$
degree gradient fluids is straightforward within the Euler-Poincar\'e
framework. However, the present paper stops at $2^{nd}$ degree gradient
fluids.

\paragraph*{Choice of energy density $W(\mathbf{e},\nabla\mathbf{e})$ for
first and second gradient fluids.}
In this paper, we will examine elliptic instability via exact nonlinear
Craik-Criminale (CC) solutions for specific cases that apply for
first and second gradient fluids. For this study, we shall choose
the strain-rate
dependence in the potential energy density as a norm,
\begin{eqnarray}\label{eq:Wform}
W(\mathbf{e},\nabla\mathbf{e})
=
\half\alpha_1|\mathbf{e}|^2
    +
\half\alpha_2|\nabla\mathbf{e}|^2,
\end{eqnarray}
where $|\mathbf{e}|^2 = e_{ij}e_{ij}$ and
$|\nabla\mathbf{e}|^2 = e_{ij,k}e_{ij,k}$ in
tensor notation. See also Bellout \etal [1999] for a discussion of
the role of this norm in proving the regularity properties of their
Rivlin-Ericksen-Green  multipolar fluids.
The case $\alpha_2 = 0, \alpha_1 \neq 0$ corresponds to the equations  for
second gradient fluids, 
\remtext{$\alpha_1 = 0, \alpha_2\neq 0$ corresponds to the
equations for third gradient fluids;}  and $\alpha_1 = 0,\alpha_2= 0$
corresponds to the classic NS equations.  For the choice in
\eref{eq:Wform}, we have
\begin{eqnarray}
\nabla\cdot{\boldsymbol\sigma}
= \alpha_1\Delta\uvec - \alpha_2\Delta^2\uvec.
\end{eqnarray}
Upon defining $\vvec = (1-\alpha_1\Delta + \alpha_2\Delta^2)\uvec$, the
motion equation (\ref{motion-eqn}) takes the following form:
\begin{eqnarray}\label{eq:uevol}
\fl \partial_t \vvec + (\uvec\cdot\nabla)\vvec + (\nabla\uvec)^T\cdot\vvec +
2\Ovec\times\uvec  - \nu\Delta\uvec \nonumber \\
\lo+
   \nabla \Big ( p - \half |\uvec|^2 - 
\half\alpha_1|\mathbf{e}|^2
    -
\half\alpha_2|\nabla\mathbf{e}|^2\Big )= \zerovec.
\end{eqnarray}
For this choice of the energy density, the stress tensor in
\eref{stress-tens} has the form:
\begin{eqnarray*}
\fl T^j_i &= m_iu_j + p\,\delta_{ij} - \left (\alpha_1e_{kj}u_{k,i} +
\half\alpha_2(e_{jk,l} + e_{kl,j})u_{k,li}\right )\, \\
\fl &= u_iu_j 
+ p\,\delta_{ij} 
- \alpha_1\left ( e_{il,l}u_j + e_{kj}u_{k,i}\right ) 
+ \alpha_2\left ( e_{il,lmm}u_j - \half(e_{jk,l} + e_{kl,j})u_{k,li}\right )\, .
\end{eqnarray*}
Note, this stress tensor is not symmetric. The Lagrangian in
\eref{Lag-Grade2-3} is also not invariant under rotations, when
the Coriolis vector potential $\Rvec(\xvec)$ is a fixed vector. In the absence
of $\Rvec(\xvec)$, this Lagrangian regains invariance under rotations and the
angular momentum in that case is conserved. However, the stress tensor
in that case is still not symmetric.

\section{CC class of solutions for gradient fluids}\label{CC-soln-sec}

A solution to \eref{eq:uevol} on an unbounded domain may be obtained,
by taking velocity in the linear form, $\uvec_0 = \Smat(t)\cdot\xvec +
\Uvec(t)$  together with a pressure $p_0$, which is quadratic in space.
The matrix $\Smat$ is a time dependent matrix 
such that
\begin{eqnarray}\label{eq:Seq}
\dot S_{ij} + S_{im}S_{mj} + 2\epsilon_{imk}\Omega_{m}S_{kj} =
M_{ij}, \qquad \qquad S_{ii} = 0,
\end{eqnarray}
and $\Uvec(t)$ is the instantaneous velocity field at the origin.
Here, $M$ is a symmetric matrix defined as $M_{ij} =
- \partial_i\partial_j\mathbb{P}$, where
\begin{eqnarray} \label{eq:bbp}
\mathbb{P} = -\int^\xvec\mathbb{F}\cdot d\xvec
+ p_0(\xvec, t) + \left (\dot \Uvec + \Smat\cdot\Uvec +
2\Ovec\times\Uvec\right )\cdot\xvec.
\end{eqnarray}
A typical approach is to choose a matrix $\Smat$ for which the left
hand side of \eref{eq:Seq} is symmetric.  Then, the corresponding
pressure $p_0(\xvec, t)$ is determined {\it a
posteriori} by \eref{eq:bbp}.
We nondimensionalize the system using the variables $\xvec' = \xvec/l$,
$t' = \omega t$, $\uvec' = \uvec/|\omega|l$, $\vvec' = \vvec/|\omega| l$,
$\alpha_1' = \alpha_1/l$, $\alpha_2' = \alpha_2/l^2$, where $l$ is a
typical length scale and
$\omega = \curl\,\uvec_0$.  The resulting equation with
the prime notation suppressed is \eref{eq:uevol} with $\nu$
replaced by $\nu/|\omega|$.

We construct a second solution to \eref{eq:uevol} of the form
$\uvec_0 + \uvec_1$ with corresponding pressure $p_0 + p_1$.  We refer
to $\uvec_0$ as the `base' flow and $\uvec_1$ as the `disturbance.'
The equations governing the disturbance are
\begin{eqnarray}\label{eq:disteqns}
\fl \partial_t\vvec_1 + \uvec_0\cdot\nabla\vvec_1 + \uvec_1\cdot\vvec_1 +
(\nabla\uvec_0)^T\cdot\vvec_1 + (\nabla\uvec_1)^T\cdot\vvec_1 +
\Pvec\times\uvec_1 \nonumber \\
+ \nabla \Big ( p_1 - \uvec_1\cdot(\nabla\cdot\mathbf{\sigma}) -
\half|\uvec_1|^2 - \alpha_1 (e_0)_{ij}(e_1 )_{ij} -
\half\alpha_1 |\mathbf{e}_1|^2 \nonumber \\
- \alpha_2 (e_0)_{ij,k}(e_1)_{ij,k} -
\half \alpha_2 |\nabla \mathbf{e}_1|^2 \Big ) = \nu\Delta\uvec_1,
\end{eqnarray}
with $\nabla\cdot\uvec_1 = 0$, 
in which we mix tensor and vector notation, where $\vvec_i =
(1-\alpha_1\Delta+\alpha_2\Delta^2)\uvec_i$ and  $\mathbf{e}_i =
\half(\nabla\uvec_i + (\nabla\uvec_i)^T)$ for $i= 0,1$, and $\Pvec =
2\Ovec + \curl\,\vvec_0$.  In the above equation, we have used the fact
that $\uvec_0$ is an exact solution to \eref{eq:uevol} together
with the vector identity
\begin{eqnarray}
\uvec\cdot\nabla\vvec + (\nabla\uvec)^T\cdot\vvec = \curl\,(\vvec)\times\uvec +
\nabla(\uvec\cdot\vvec)
\end{eqnarray}
for any two vectors $\uvec,\vvec$.  
We choose the disturbance to be of the form 
\begin{eqnarray}
&&\uvec_1 = \mu \avec(t)\sin(\beta\psi(\xvec,t)), \\
&&        p_1 = \mu \hat p_{11}(t)\cos(\beta\psi(\xvec,t))
          + \mu^2\hat p_{12}(t)\cos^2(\beta\psi(\xvec,t)),
\end{eqnarray}
$\psi(\xvec,t) = \dotp{\kvec(t)}{\xvec} + \delta(t)$, and
$\mu$ and $\beta$ are scaling factors so that we can choose the initial
conditions $|\avec(0)| = 1$ and $|\kvec(0)| = 1$.
The unknown phase $\psi(\xvec,t)$ and the amplitudes
$\avec(t)$, $\hat p_{11}(t)$, and $\hat p_{12}(t)$ are to be determined.
The incompressibility condition
$\nabla\cdot\uvec_1 = 0$ gives
\begin{eqnarray}
\dotp{\avec}{\kvec} = 0 \label{eq:incomp}.
\end{eqnarray}
Form this equation it follows that the nonlinear term
$\uvec_1\cdot\vvec_1$ in \eref{eq:disteqns} vanishes exactly.
Thus, in what follows, the sum $\uvec_0 + \uvec_1$ is an exact
solution to the nonlinear equations of motion in \eref{eq:uevol}.
By collecting on powers of $\sin(\beta\psi)$ and $\cos(\beta\psi)$,
the evolution equations for the amplitudes and phase are
\begin{eqnarray}
&&p_{12} - (\Upsilon-1)|\avec|^2 +
\half \beta^2|\avec|^2|\kvec|^2(\alpha_1-\alpha_2|\kvec|^2\beta^2) =
0,\label{eq:p2evol} \\
&&\partial_t\psi + (\Smat\cdot\xvec +\Uvec)\cdot\kvec = 0, \label{eq:xievol} \\
&&d_t{\Big ( \Upsilon\avec\Big )}
          + \Upsilon\Smat^T\cdot\avec
          + \Pvec\times\avec
          - \beta\tilde P\kvec =
          - E_\omega|\kvec|^2\avec .
   \label{eq:aevol}
\end{eqnarray}
Here ,
\begin{eqnarray}
\Upsilon(t) = 1+\alpha_1\beta^2|\kvec(t)|^2+\alpha_2\beta^4|\kvec(t)|^4,
\end{eqnarray}
$E_\omega =\nu\beta^2/|\omega|$ is the vorticity based Ekman number,
$\Pvec = \curl\,\uvec_0 + 2\Ovec$ is the total vorticity of the
system, and $\tilde P =  p_{11} -
\half\beta^2\alpha_1\avec\cdot(\Smat+\Smat^T)\cdot\kvec$.
Note that the amplitude scaling $\mu$ is
immaterial.
Without loss of generality, we set
\begin{eqnarray}\label{eq:deltavan}
d_t \delta + \dotp{\kvec}{\Uvec} = 0.
\end{eqnarray}
Then taking the gradient of \eref{eq:xievol} becomes
\begin{eqnarray}
d_t{\kvec} + \Smat^T \cdot \kvec &= 0. \label{eq:kevol}
\end{eqnarray}
We eliminate the pressure term by
taking the dot product of \eref{eq:aevol} with $\kvec$ and by using
   $\dotp{{\mathrm d}\avec/{\mathrm dt}}{\kvec} = - \dotp{\avec}{
{\mathrm d}\kvec/{\mathrm dt}} =\dotp{(\Smat\cdot\avec)}{\kvec}$, the
first of which follows from \eref{eq:incomp} and the second from
\eref{eq:kevol}:
\begin{eqnarray}\label{eq:pressure}
\beta\tilde P =
          \frac{1}{|\kvec|^2}\Big \{ \Upsilon\dotp{[(\Smat+\Smat^T)\cdot\avec]}{\kvec}
          + \dotp{\Pvec\times\avec}{\kvec} \Big \}.
\end{eqnarray}
In summary, we have obtained a new exact incompressible solution to
\eref{eq:uevol}.
The variables are amplitude $\avec(t)$ and wave vector $\kvec(t)$.  Once
these are determined, the pressure terms follow from \eref{eq:p2evol} and
\eref{eq:pressure}.
Note that $\uvec_0$ and $\uvec_0 + \uvec_1$
are exact solutions to the nonlinear equations,
but $\uvec_1$ by itself is only a solution to \eref{eq:uevol} linearized
about $\uvec_0$.  The exception is that in a rotating coordinate
system ($\Ovec \neq \zerovec$),
$\uvec_1$ is an
exact solution by itself since this scenario corresponds to
$\uvec_0 = \mathcal{R}\cdot\xvec$ in a non-rotating frame, where
$\mathcal{R}$ is rigid body
rotation about the $z-$axis;
cf. Ref.~\cite{lif:miya:fab:98}.
We emphasize that the operator
$d_t + S^T$ acting on a vector represents the
complete time
derivative of that quantity in a Lagrangian frame moving with $\uvec_0$.
Finally, the equation for $\avec(t)$ is\footnote[5]{
Alternatively, one can collect on the terms linear and constant in
$\xvec$ upon insertion of $\uvec_0+\uvec_1$ into \eref{eq:uevol}.  In
either case, by enforcing \eref{eq:deltavan}, both methods yield the
same set of equations: \eref{eq:p2evol},\eref{eq:kevol}-\eref{eq:aevol2}.}
\begin{eqnarray}\label{eq:aevol2}
\fl d_t\avec = \frac{1}{\Upsilon} \Big \{ 2\beta^2(\Smat\cdot\kvec)\cdot\kvec (\alpha_1
+ 2\alpha_2\beta^2|\kvec|^2)\avec - \Upsilon\Smat^T\cdot\avec -
\Pvec\times\avec +
\tilde P\kvec - E_\omega|\kvec|^2\avec \Big \}.
\end{eqnarray}

\section{Elliptic instability for gradient fluids}
\label{Elliptic-Instab-sec}

We examine the stability of a rotating column of fluid with
elliptic streamlines whose foci lie on the $y$-axis:
\begin{eqnarray}
\uvec_0 = \half\omega L\cdot\xvec, \quad L = \pmatrix{
   0 & -1+\gamma & 0 \cr 1+\gamma & 0 & 0 \cr
          0 & 0 & 0 }.
\end{eqnarray}
Here, $0 \leq \gamma < 1$ is the eccentricity of the ellipses, and the pressure
is $p_0 = \frac{1}{2}\omega^2 (1-\gamma^2)(x^2+y^2)$.
\Eref{eq:kevol} with $\Smat=L$ is analytically solvable:
\begin{eqnarray} \label{eq:ksol}
\kvec = [\sin\theta\cos(t\sqrt{1-\gamma^2}),
          \kappa\sin\theta\sin(t\sqrt{1-\gamma^2}), \cos\theta]^T
\end{eqnarray}
where $\kappa^2 = (1-\gamma)/(1+\gamma)$ and $\theta$ is
the polar angle that $\kvec$ makes with the
axis of rotation.
\Eref{eq:aevol2} has the
form
\begin{eqnarray*}
d_t\avec = \Nmat(t;\alpha_1,\alpha_2,E_\omega,\Omega,\gamma,\theta)\cdot \avec,
\end{eqnarray*}
where the
elements of the matrix $\Nmat$ are periodic with period
$\tau = 2\pi/\sqrt{1-\gamma^2}$, the period of $\kvec(t)$.  Therefore,
the system can be analyzed
numerically using Floquet theory \cite{yaku:star:76}.  We compute the
monodromy matrix $\Pmat$, that is, the fundamental solution matrix
with identity initial condition evaluated at $t = \tau$.  
\Eref{eq:aevol} will have exponentially growing solutions if
$\max_i|\Re (\rho_i)| > 1$, where $\rho_i, i=1,2,3$ are the eigenvalues
of $\Pmat$, with corresponding Lyapunov-like growth rates given by
\begin{eqnarray*}
\sigma = \ln \{\max_i|\Re (\rho_i)|\}/\tau.
\end{eqnarray*}
Thus, we can simulate numerically the solution to \eref{eq:aevol}
over one period and indisputably determine the exponential
growth rates.
We can be certain that at least one of the eigenvalues will always
be unity because
of the incompressibility condition \eref{eq:incomp} and
that the remaining two eigenvalues
appear as complex conjugates on the unit circle or as real valued
reciprocals of each other.

\subsection{Inviscid results for gradient fluids}
For flows with circular streamlines ($\gamma = 0$),
the monodromy matrix can be analytically computed.
It follows from \eref{eq:ksol}
that $|\kvec(t)| = 1$.
Then, $\Upsilon$
is constant in time (denoted by $\Upsilon_0 =
1+\alpha_1\beta^2+\alpha_2\beta^4$)  and
\eref{eq:aevol} has
three linearly independent solutions:
\begin{eqnarray}
\avec_1(t) &= \cos(\xi(t)+\phi) \kvec_{\perp 1} + \sin(\xi(t)+\phi)
\kvec_{\perp 2} \label{eq:a1} \\
\avec_2(t) &= \sin(\xi(t)+\phi) \kvec_{\perp 1} - \cos(\xi(t)+\phi)
\kvec_{\perp 2} \\
\avec_3(t) &= \ez,
\end{eqnarray}
where $\xi(t) = 2t(1+\Omega)\cos\theta /\Upsilon_0$,
$\kvec_{\perp 1} = [\cos\theta \cos t, \cos\theta \sin t,
-\sin\theta]^T$ and $\kvec_{\perp 2} = [\sin t, -\cos t, 0]^T$ are
vectors orthogonal to $\kvec$, and $\phi$ is an arbitrary phase.
Clearly the first two solutions $\avec_1$ and $\avec_2$
satisfy \eref{eq:incomp}.  The monodromy matrix can be constructed from
these three solutions:
\begin{eqnarray*}
\Pmat = \pmatrix{
\cos(\xi(2\pi)) & \cos\theta\sin(\xi(2\pi)) & 0 \cr
-\sin(\xi(2\pi))/\cos\theta & \cos(\xi(2\pi)) & 0 \cr
\tan\theta (1-\cos(\xi(2\pi))) & -\sin\theta\sin(\xi(2\pi)) & 1
}.
\end{eqnarray*}
The three eigenvalues are
$\rho_{1,2} = \exp(\pm i\xi(2\pi)), \rho_3 = 1$.
All of the eigenvalues lie on the unit circle,
from which it follows that all solutions
in the inviscid case for $\gamma = 0$ are stable.  The values
of $\cos\theta$ for which $|\rho_i| = 1, i=1,2,3$ are called `critically
stable' and are given by $\xi(2\pi) = m\pi$, $m = 0, \pm 1, \pm 2, \ldots$.
At these parameter values an exponentially growing solution can
appear (together with an exponentially decaying one)
as $\gamma$ increases from zero.
Bayly \cite{bayly:86} argues that the evenness of
$\tilde P\kvec$ as a function of $\kvec$ implies that the
eigenvalues, if real and unequal, must be positive.  This dismisses
the odd choices of $m$.  Furthermore, Floquet theory is not
applicable for the case $m=0$.  Thus, the possible choices for
critical stability are $\xi(2\pi) = 2n\pi$, $n=\pm 1, \pm2, \ldots$.
This corresponds to
\begin{eqnarray}\label{eq:critangle}
\cos\theta = \frac{n\Upsilon_0}{2(1+\Omega)}.
\end{eqnarray}
These are the critical parameter value at which $\avec(t)$ suffers exponential
growth as $\gamma$ increases from zero.  For the NS equations
(i.e. $\Upsilon_0 = 1$), only the $n=1$ choice (called the `principle
finger') is physically
interesting.  The other choices of $n$ are extremely thin fingers with
growth rates ten orders of magnitude smaller than that of the
principle finger \cite{fab:holm:03c}.  As $\alpha_1$ and/or $\alpha_2$
increase from zero,
however, the fingers widen and the associated growth rate increases.
Finally, since $|\cos\theta|\leq 1$, we conclude that there exists a
band of stable eccentricities for
\begin{eqnarray}\label{eq:olims}
-\frac{\Upsilon_0}{2} < \Omega + 1 < \frac{\Upsilon_0}{2}.
\end{eqnarray}

Additional understanding of this result emerges by following the
analysis of
Waleffe \cite{wal:90} and Kerswell \cite{kers:02}.
By taking the dot product of \eref{eq:aevol} with $\avec$, we
obtain (for all $\gamma$ and $\Upsilon$)
\begin{eqnarray}
\fd{\left ( \half |\avec|^2 \right ) } = -2\gamma a_1 a_2 +
\frac{4\gamma(\Upsilon-1)}{\Upsilon}\frac{k_1k_2}{|\kvec|^2}|\avec|^2.
\end{eqnarray}
One can determine an exponential growth rate to leading order in
$\gamma$ by inserting the zeroth order solutions for $\kvec$ and
$\avec_1$ into the right hand side of this equation:
\begin{eqnarray}
 \sigma &\equiv \frac{1}{|\avec|^2}\fd{\left ( \half|\avec|^2\right
) }\nonumber \\ 
 &= -\frac{\gamma}{4}[(1-\cos\theta)^2\sin(2(\xi_++\phi)) 
   - (1+\cos\theta)^2\sin(2(\xi_-+\phi)) \nonumber \\
&\quad    - 2(1-\cos^2\theta)\sin(2t)]
   + \frac{2\gamma(\Upsilon_0-1)}{\Upsilon_0}\sin^2\theta\sin(2t),
\end{eqnarray}
where $\xi_\pm = \xi(t) \pm t$.  Upon averaging over a period of
$\avec_1$, this quantity will vanish
except when $\xi_\pm = 0$, corresponding to $\cos\theta =
\mp\Upsilon_0/2(1+\Omega)$.  Compare this with
\eref{eq:critangle}.  The
maximum values for $\sigma$ will occur at $\phi = \mp\pi/4$ for
$\xi_\pm = 0$, respectively, with growth rate
\begin{eqnarray}\label{eq:grrate}
\sigma_\mathrm{max} =
\frac{(2+\Upsilon_0)^2}{16}\times\frac{(2+\Upsilon_0+2\Omega)^2}{(2+\Upsilon_0)^2(1+\Omega)^2}\gamma 
+ O\Big(\gamma^2\Big ),
\end{eqnarray}
valid for $\Upsilon_0 \leq 2$ and $\Omega$ not satisfying
\eref{eq:olims}.  Thus, we
see that the maximum growth rate increases as a function of
$\alpha_1$ and $\alpha_2$ due to the $\Upsilon_0$ dependence of the
critical stability point
up to a maximum of $\sigma = \gamma$,
after which a set of stable solutions emerges in a band of nonzero
eccentricities.    See Fig.~\ref{fig:sample}
\remfigure{
\begin{figure}
\begin{center}
\includegraphics[width=5in]{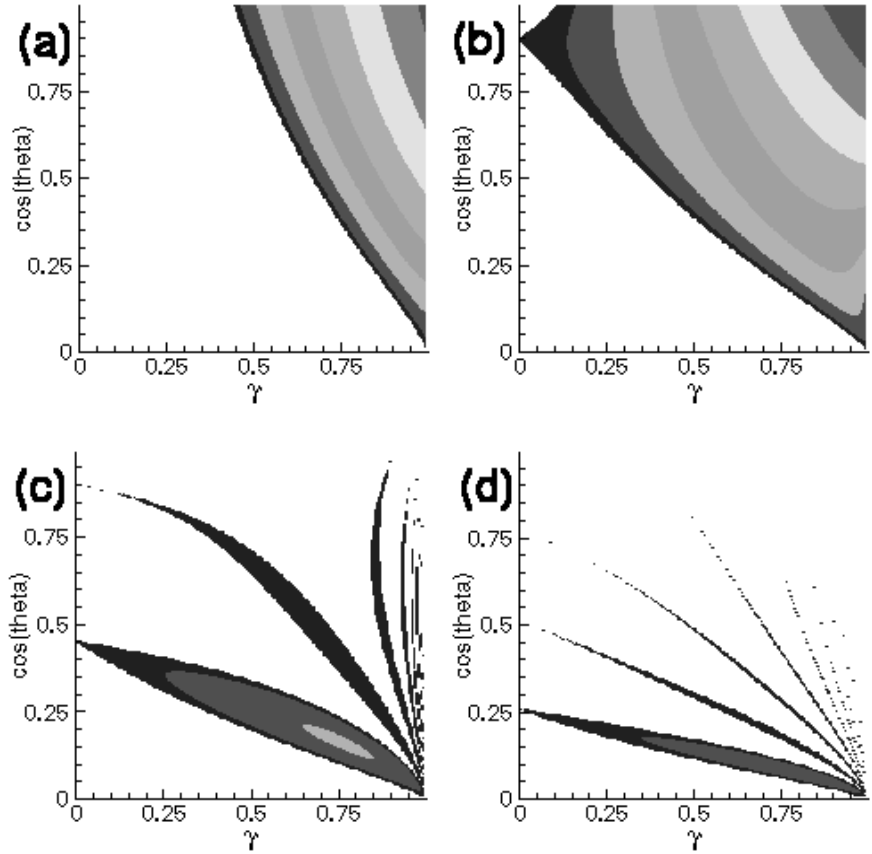}
\caption{Sample contour plots of instability regimes for $\alpha_1 =
0.5$, $\alpha_2 = 0.3$, $\beta = 1.0$, $E_\omega = 0$ computed on a
$250\times 250$ grid for various
values of $\Omega$:  (a) -0.5, (b) 0, (c) 1.0, and (d) 2.5.  Note that
the individual fingers touch the $\cos\theta$ axis according to
\eref{eq:critangle}.
\label{fig:sample}
}
\end{center}
\end{figure}
}

For nonzero values of $\gamma$, we must investigate the system
numerically.
We use the variable coefficient ordinary differential equation 
solver \textsf{DVODE}
\cite{dvode}. The level surface of the growth rate for fixed $\alpha_1 = 0$
is seen in Fig.~\ref{fig:bigfigs1}, and Fig.~\ref{fig:a1a2} shows the
growth rate surface maximized over the $\gamma,\cos\theta$ plane as a
function of $\alpha_1,\alpha_2$.  \newt{Numerical
experiments show that $\sigma_\mathrm{max}$ has the value associated 
with the NS equations for 
$\alpha_1=\alpha_2 = 0$.  As the parameters
$\alpha_{1,2}$ increase, $\sigma_\mathrm{max}$ increases to a value of
unity on the line
$\alpha_1\beta^2+\alpha_2\beta^4=1+2\Omega$, and then decreases slowly to zero
as $\alpha_{1,2}\to\infty$.  This threshold line corresponds to 
the maximal rate of change of $\sigma_\mathrm{max}$ 
in \eref{eq:grrate} with respect to $\gamma$.  See Fig.~3.}
\remfigure{
\begin{figure}
\begin{center}
\includegraphics[width=2.5in]{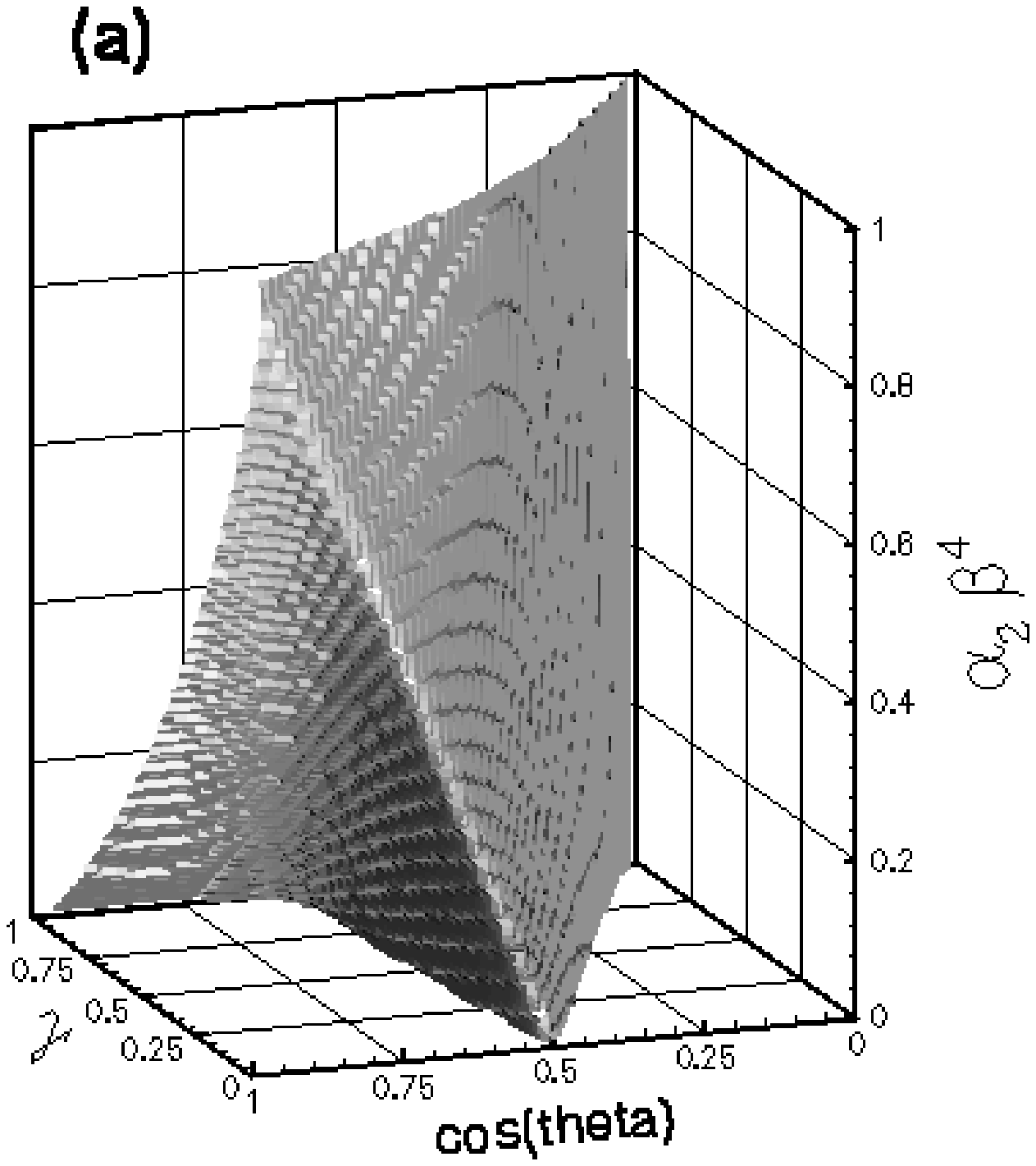}
\includegraphics[width=2.5in]{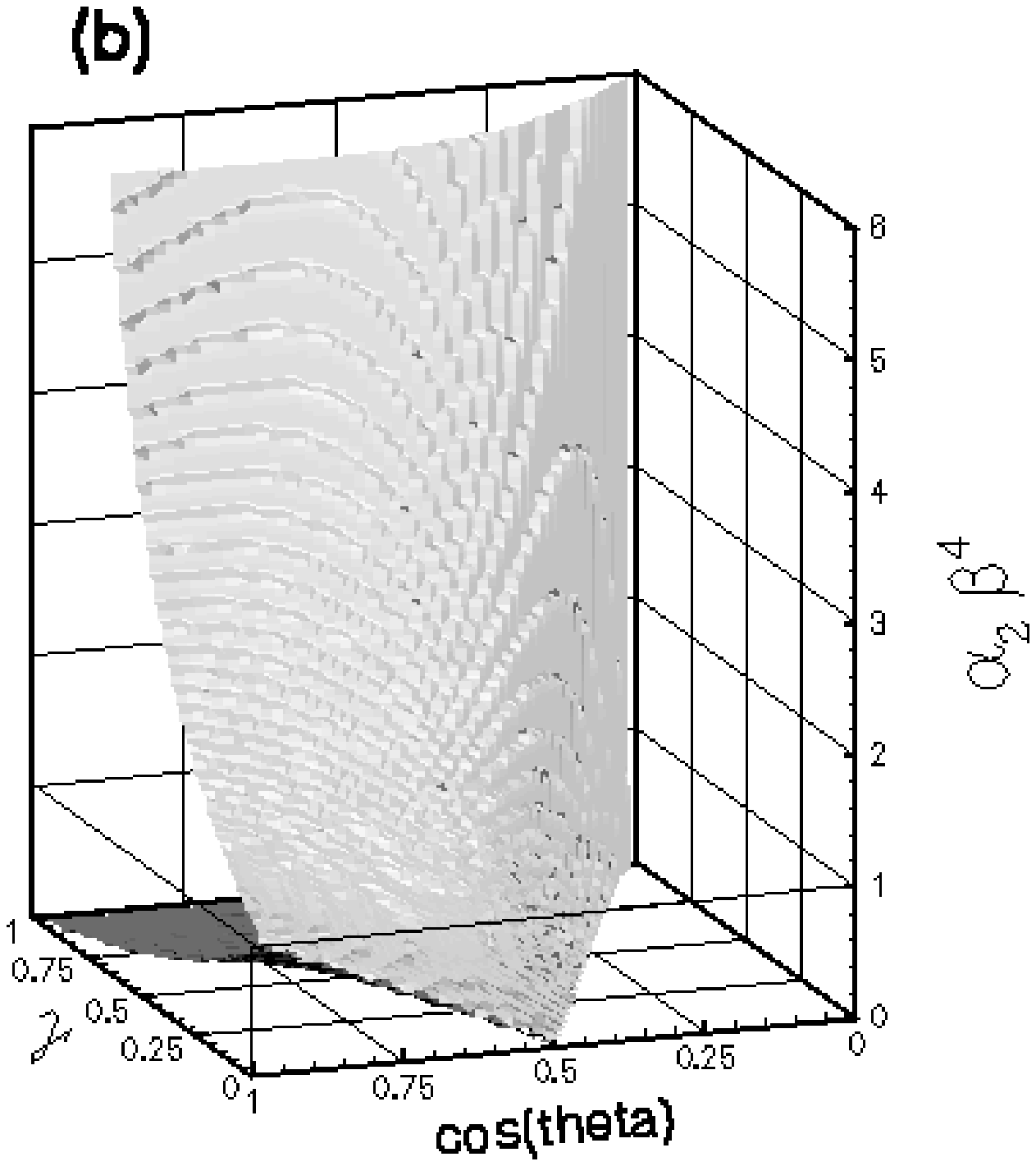}
\caption{Surface of $\sigma = 0.01$ for $E_\omega = 0$, $\alpha_1 =
0$, $\beta = \sqrt{2}$, $\Omega = 0$, and various $\alpha_2$.
Figure (a) shows the neutral surface for $0 \leq \alpha_2 \leq 0.25$ and
is an expansion of the boxed region in figure (b).
For $\alpha_2 = 0$, the critical stability point
occurs at $\theta = \pi/3$, which agrees with the classical NS results.
The critical stability point shifts towards $\cos\theta = 1$
as $\alpha_2$
increases according to $\cos\theta = (1+\alpha_2\beta^4)/2$.
As $\alpha_2\beta^4$ exceeds
unity, a stable band of rotating flows with nonzero
eccentricities appears.  The corresponding surfaces for $\alpha_2 = 0$
and various $\alpha_1$ are qualitatively the same, \newt{albeit as a
function of $\alpha_1\beta^2$}.  See \cite{fab:holm:03a}.
\label{fig:bigfigs1}
}
\end{center}
\end{figure}
}
\remfigure{
\begin{figure}
\begin{center}
\includegraphics[width=2in]{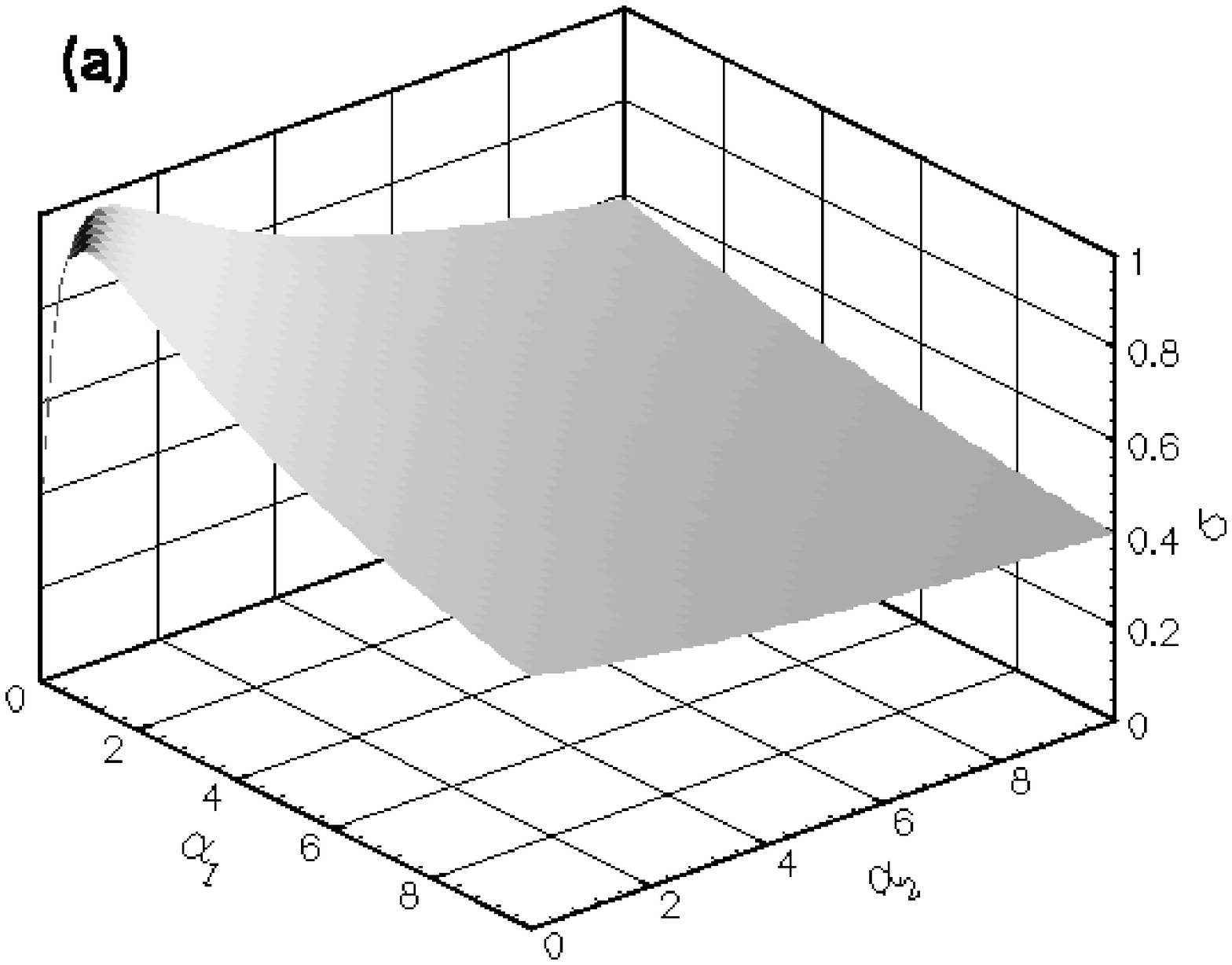}
\includegraphics[width=2in]{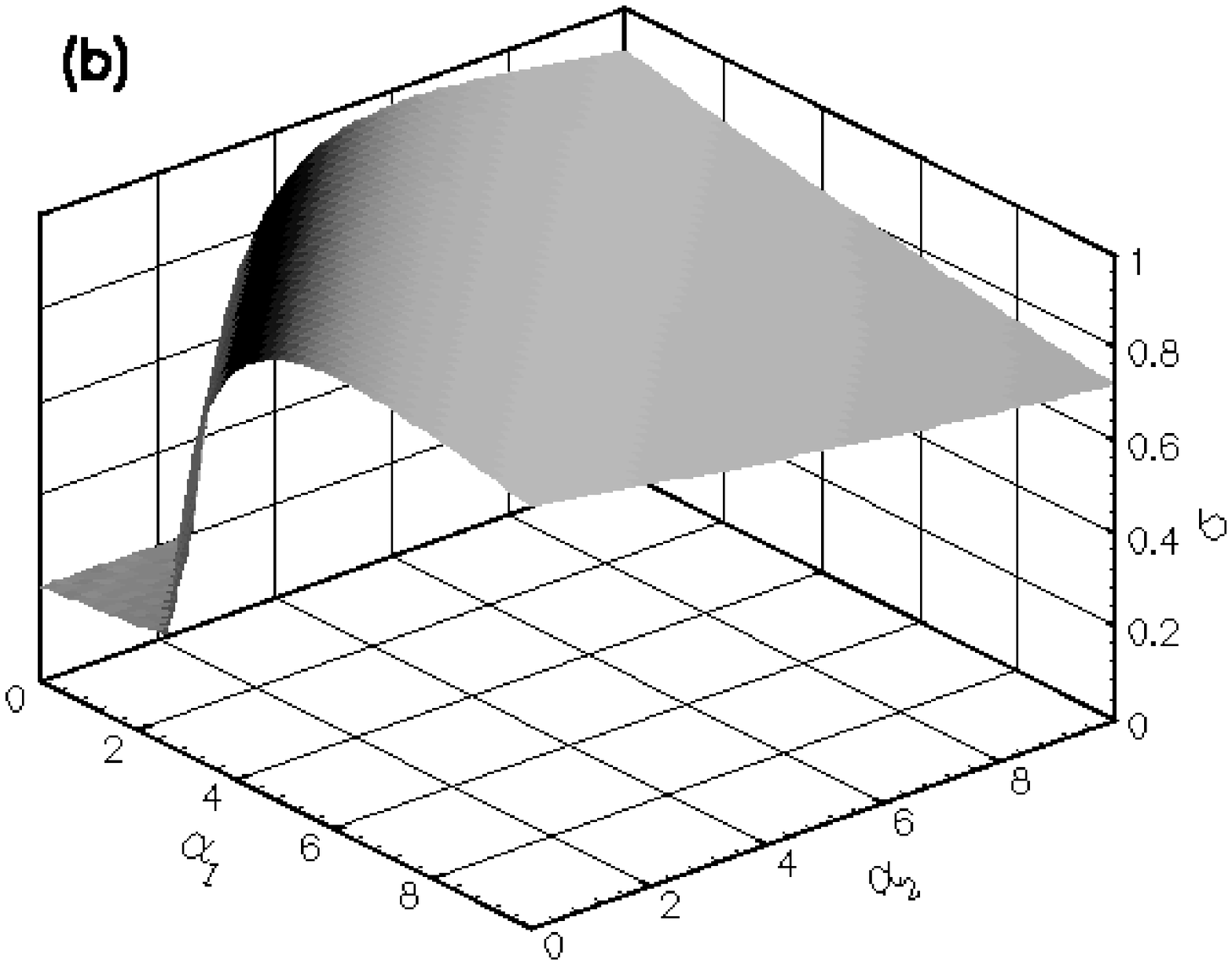}
\caption{Growth rate surface $\sigma$ maximized over the $\cos\theta,\gamma$
plane as a function of $\alpha_1$ and $\alpha_2$ for $E_\omega = 0$,
$\beta = 1$, (a) $\Omega = 0$
and (b) $\Omega = 2.5$.  In figure (a), the growth rate for
$\alpha_1=\alpha_2=0$ (corresponding to the classic NS case) is
$0.36$.  We see that the growth rate quickly increases to unity
\newt{on the line $\alpha_1\beta^2+\alpha_2\beta^4 = 1+2\Omega$} and
then slowly decays.  Not shown is that $\sigma\to 0$ as
$\alpha_1,\alpha_2\to\infty$.
\label{fig:a1a2}
}
\end{center}
\end{figure}
}

\subsection{Viscous results for gradient fluids}
The solutions to \eref{eq:aevol2} must be simulated numerically
for $E_\omega \neq 0$.  An interesting feature of this equation is
that, unlike the NS equations, a change of variables will not remove
viscosity from the problem.  However, the qualitative results for NS
hold true here.  Viscosity stabilizes the flow by lowering the maximum
growth rate and introducing a stable band of eccentricities.  This
stabilization is slower than its NS counterpart, that is, 
the disspation in \eref{eq:uevol} is of the form $\nu\Delta\uvec$, not
$\nu\Delta\vvec$.  See Figs.~\ref{fig:viscfixedbeta} and
\ref{fig:visfixedalpha}.
\remfigure{
\begin{figure}
\begin{center}
\includegraphics[width=2in]{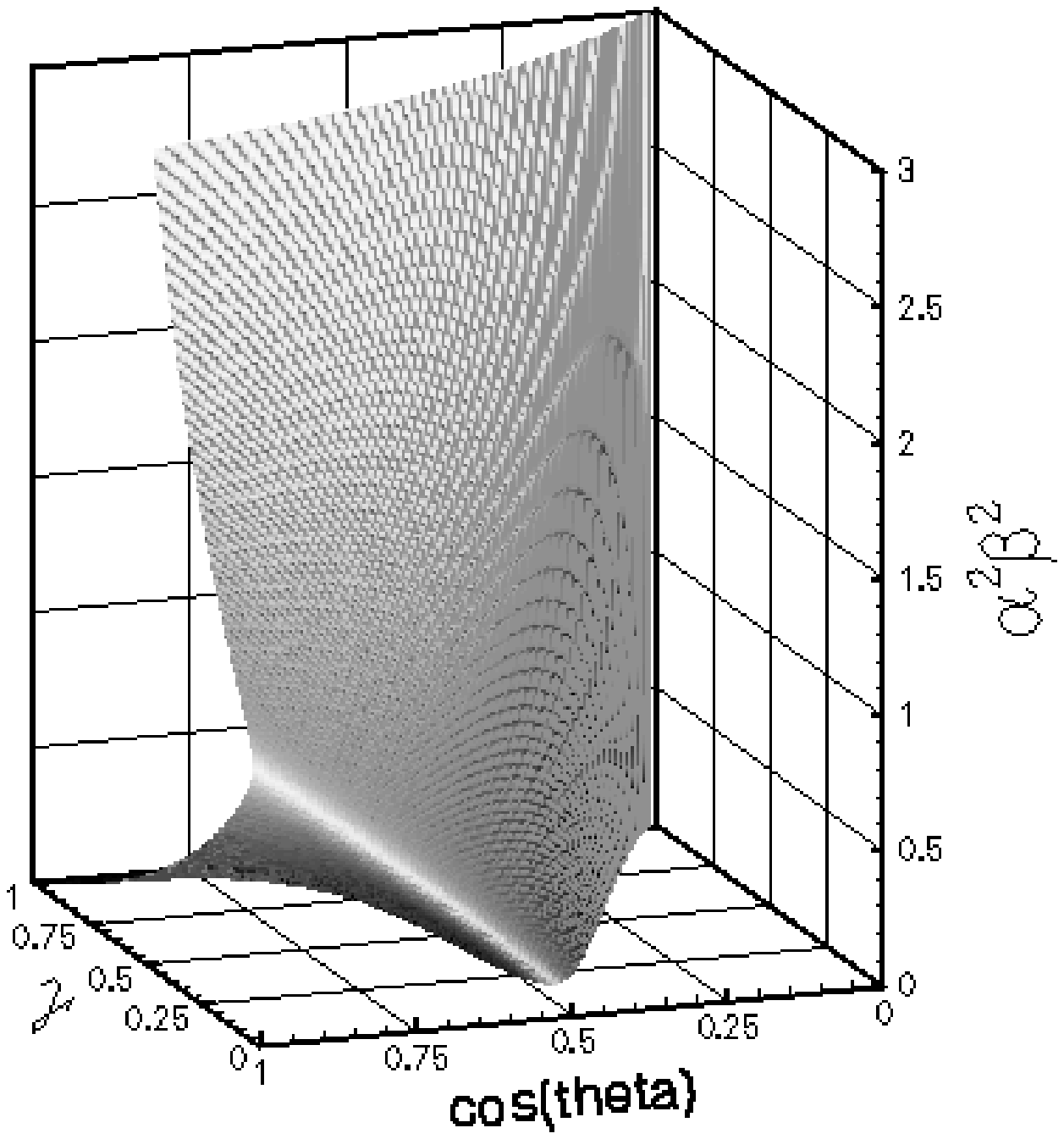}
\includegraphics[width=2in]{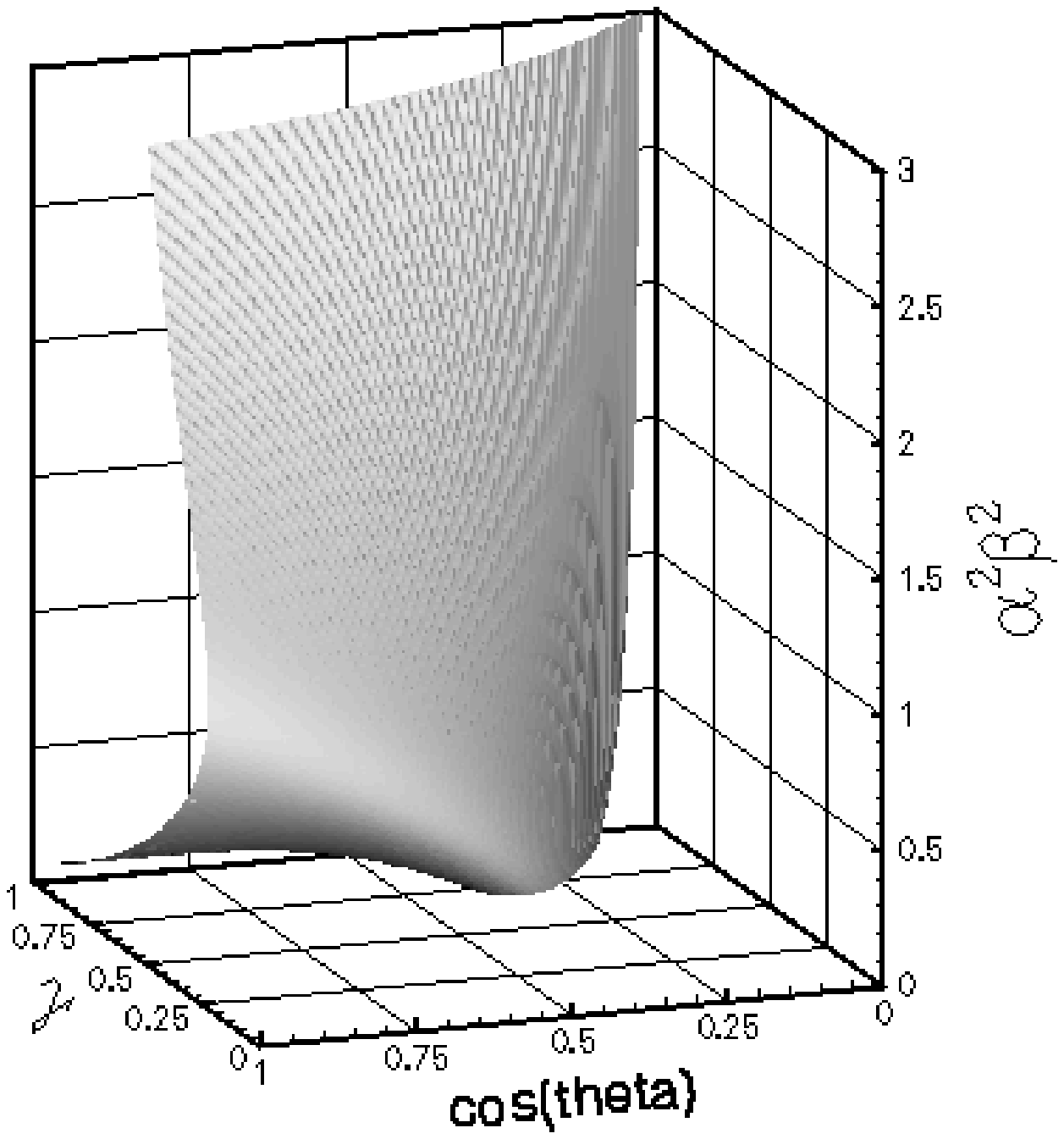}
\includegraphics[width=2in]{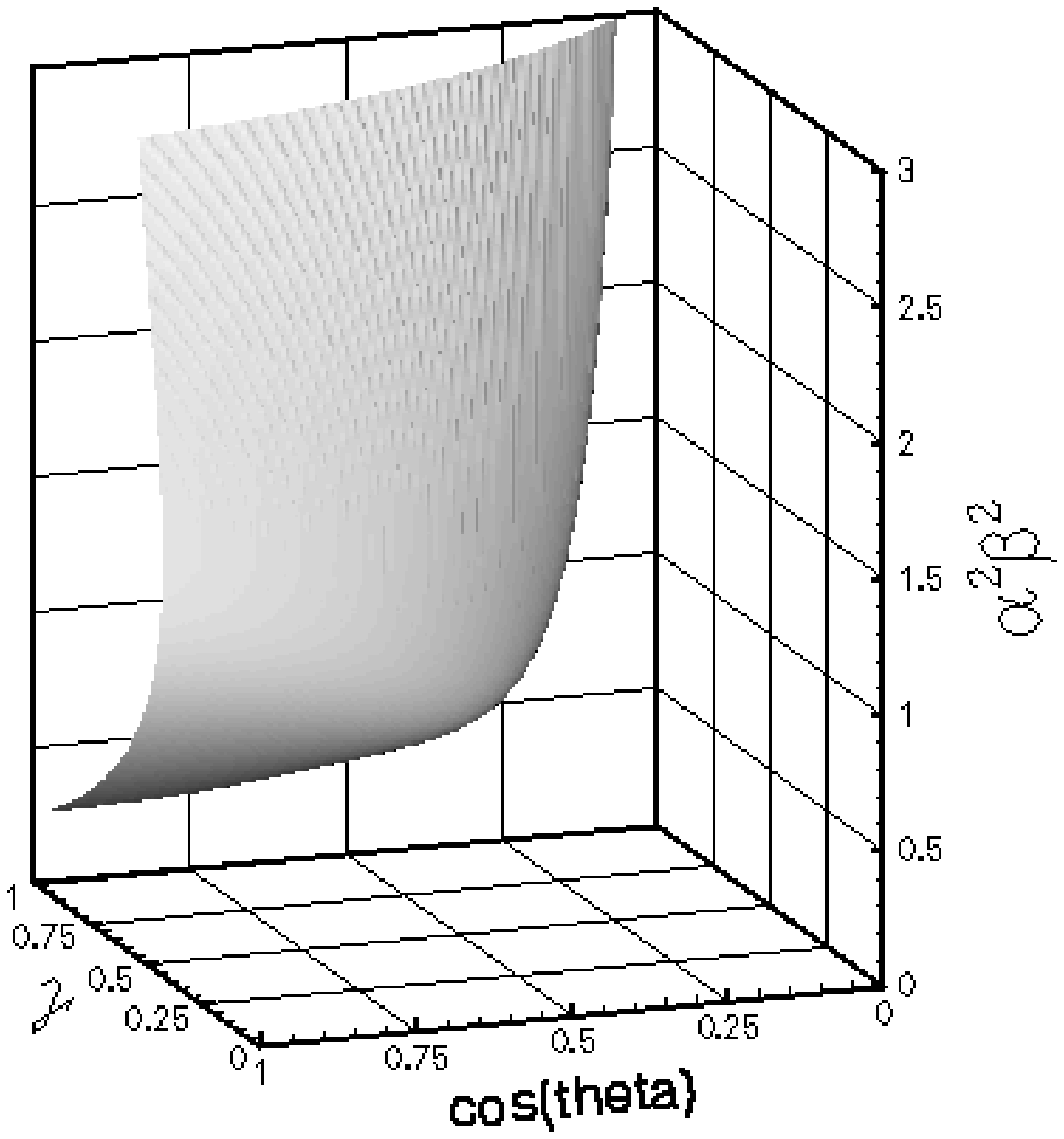}
\caption{Surface of $\sigma = 0.01$ for $\alpha_2 =
0$, $\beta = 1$, $\Omega = 0$, as functions of  $\alpha_1$ for (a)
$E_\omega = 0.1$, (b) $E_\omega = 0.5$, and (c) $E_\omega = 1.0$, 
\newt{$E_\omega = \nu\beta^2/|\omega|$}.
Again, the corresponding surfaces for $\alpha_1 = 0$
and various $\alpha_2$ is qualitatively the same.
\newt{Namely, the unstable region shrinks as viscosity in
  $E_\omega$ increases.}  
\label{fig:viscfixedbeta}
}
\end{center}
\end{figure}
}
\remfigure{
\begin{figure}
\begin{center}
\includegraphics[width=2.5in]{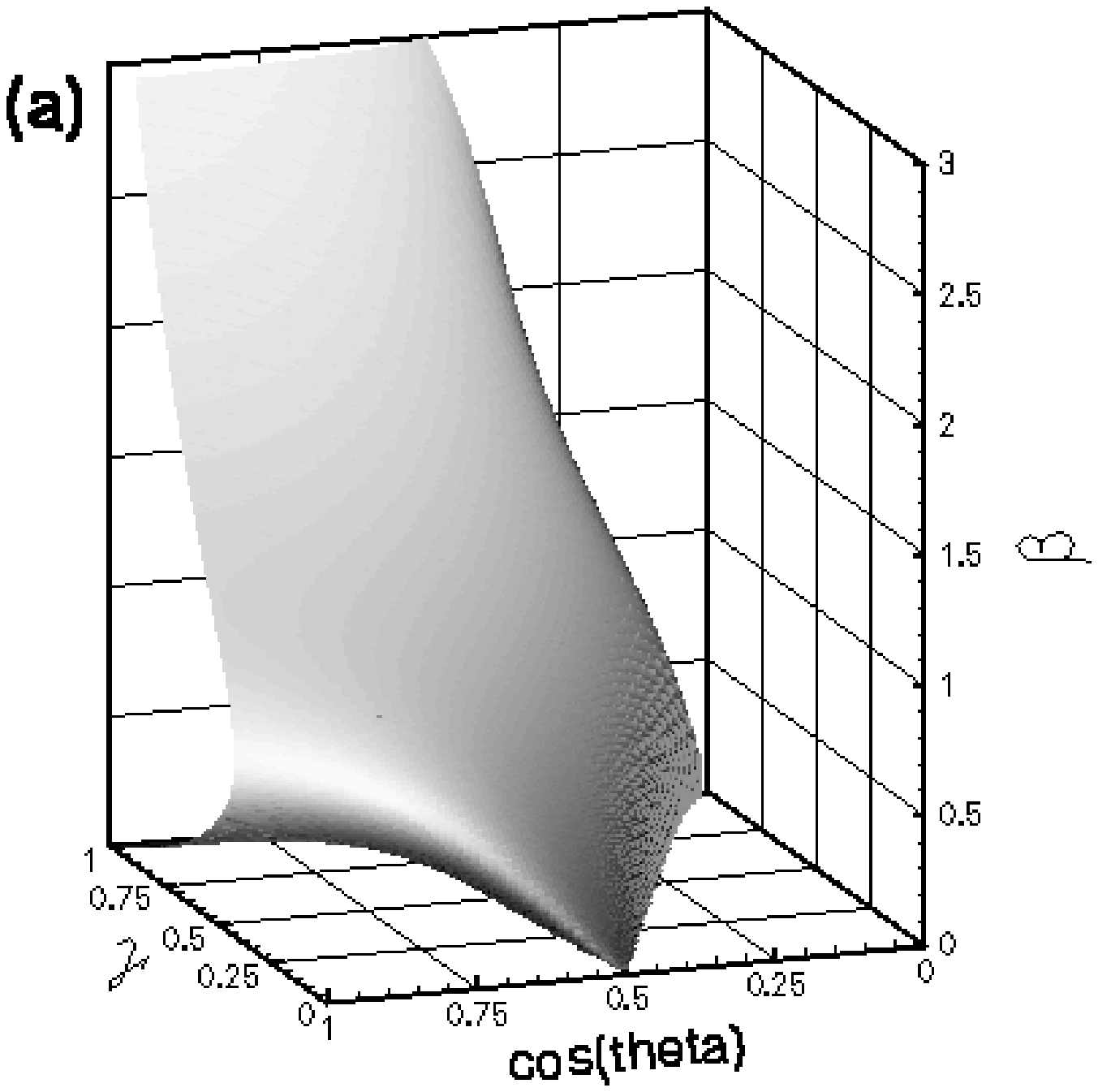}
\includegraphics[width=2.5in]{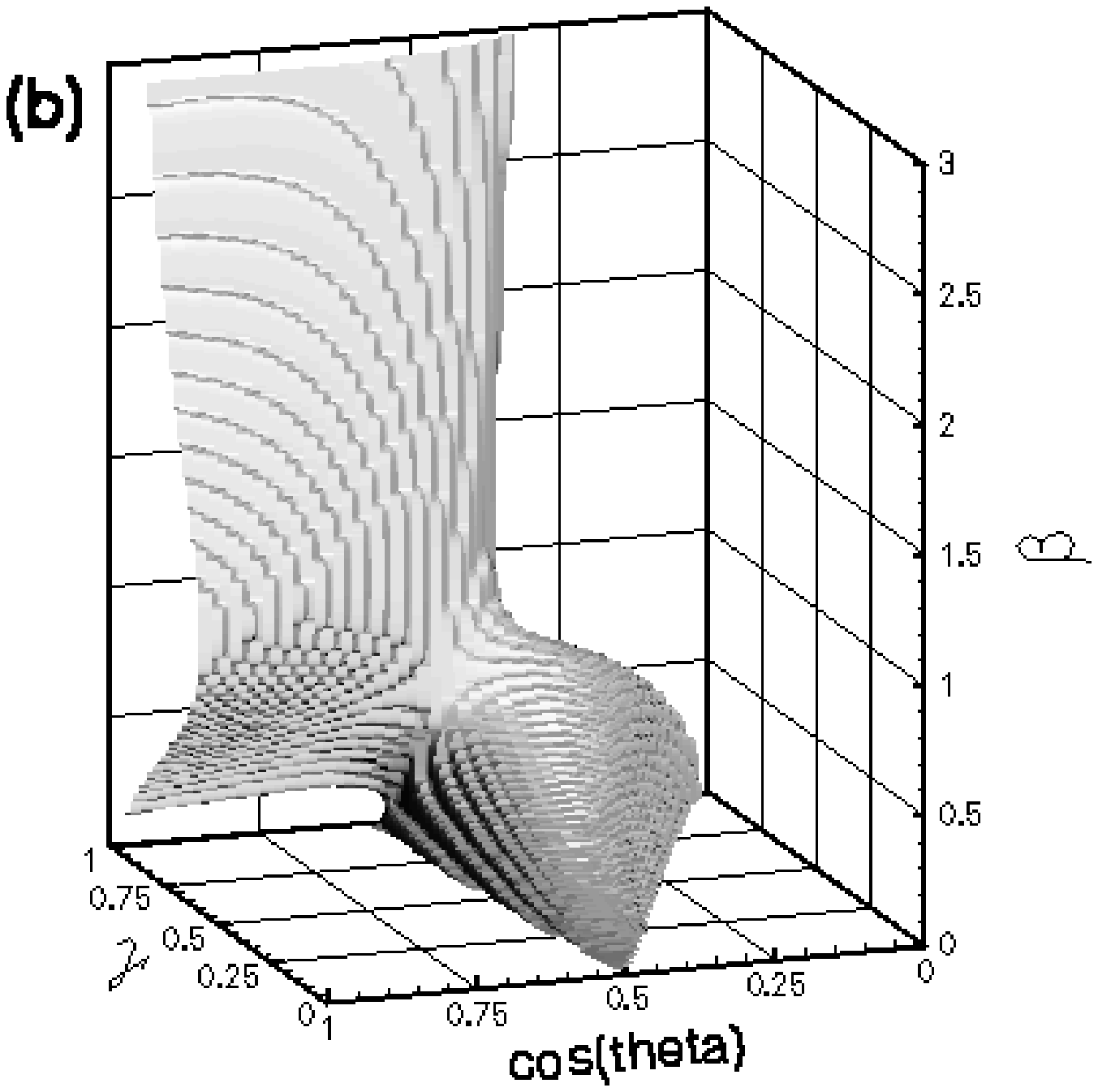}
\caption{Surface of $\sigma = 0.01$ for $\Omega = 0$, $\nu/|\omega| =
1$ as a function of
the wavenumber $\beta$ for (a) $\alpha_2 =
0$, $\alpha_1 = 1$ and (b) $\alpha_1 = 0$, $\alpha_2 = 1$.   For
$\beta \ll 1$, the flow is inviscid.  As
$\beta$ increases, the leading order term in an asymptotic exapnsion
of \eref{eq:aevol} will be $-E_\omega|\kvec|^2\avec$.  Since
$E_\omega = \nu\beta^2/|\omega|$, viscous dissipation quickly takes over.
\label{fig:visfixedalpha}
}
\end{center}
\end{figure}
}

\section{Conclusions}\label{conclusion-sec}
\newt{%
The presence of nonlinear elasticity was seen to have profound effects
on the properties of elliptic instability.  It can affect the
growth rates, as well as the shapes and sizes of the 
unstable parameter regimes.  One of the most
profound effects is the thickening of the resonance domains (fingers,
or Arnold tongues) in Fig.~1.  These resonance domains of instability
were predicted for the NS elliptic instability.  However, in the NS
case, they are infinitesimally thin.}

\newt{%
The second gradient fluid
constituitive relation  and the LANS$-\alpha$ turbulence model 
both introduce higher derivatives in the
momentum density.  We found that 
the highest derivative dominates and produces qualitatively
similar effects on the neutral stability surfaces.  That is, Fig.~2
shows a similar behavior of the neutral surface
as a function of $\alpha_2\beta^4$ as found
for the LANS$-\alpha$ model as a function of $\alpha_1\beta^2$, in the
present notation.}

\newt{%
As seen in Fig.~3, first and second
gradient fluids {\it increase} the Lyapunov growth rates associated
with elliptic instability for $\alpha_1\beta^2+\alpha_2\beta^4 < 1+2\Omega$
and  then {\it decrease} the growth rates for parameter values beyond
this threshold.
When $\alpha_2 = 0$, this relation recovers the result for
LANS$-\alpha$.  Thus, the higher-order smoothing due to $\alpha_2\neq
0$ comes into play to reduce the maximum growth rate for short waves.}

\newt{%
Viscosity has the expected effects on the domain
of  elliptic instability, as seen in Figs.~4 and 5.  
However, these effects depend sensitively
on the value of $\alpha_1$ and $\alpha_2$.  Figure~5 shows how the
effects of nonlinear {\it visco}elasticity depend on the values of
$\alpha_1$ and $\alpha_2$ as a function of the wave number $\beta$.
The $\alpha_2$ term corresponds to the $\beta^4$ dependence, which
comes into play very rapidly in its effect on the neutral surface for
elliptic instability in Fig.~5b.  }  

\newt{%
Our investigation followed the approach of Fabijonas and Holm
\cite{fab:holm:03a,fab:holm:03c}, who 
studied the corresponding mean effects of turbulence on elliptic
instability for a class of turbulence closure
models. For inviscid fluids, the 
effects of elliptic instability seen in gradient
fluids and in the turbulence closure models are qualitatively similar.
The inviscid first gradient fluid corresponds to the LANS$-\alpha$
turbulence model, which can be viewed as the nonlinear terms 
in an LES model for turbulence whose
filter is the inverse of the Helmholtz operator
$(1-\alpha_1\nabla^2)$ \cite{foias:holm:titi:01}.  
The inviscid second gradient
fluid can be viewed similarly, for which the filter is
$(1-\alpha_1\nabla^2+\alpha_2\nabla^4)^{-1}$, instead.  }

Future studies may investigate the roles of other aspects of nonlinear
stress on elliptic instability, for example, in the
Rivlin-Ericksen-Green multipolar fluids analyzed in Bellout \etal
\cite{Bellout-etal[1999]}.

\ack
BRF thanks the Los Alamos National Laboratory for its hospitality, 
and gratefully acknowledges the financial
support of the Laboratory's Turbulence Working Group and  
Theoretical Division.  DDH is grateful for partial support by Imperial
College London and by US DOE, under contract W-7405-ENG-36 for Los
Alamos National Laboratory, and Office of Science ASCAR/AMS/MICS.


\end{document}